\documentclass[final,english]{bullsrsl}[2022/06/15]

\usepackage[latin1]{inputenc}
\usepackage[T1]{fontenc}

\usepackage{aas_macros}
\usepackage{natbib} 
\usepackage{graphicx}
\usepackage{hyperref}
\usepackage{amsmath, amssymb, siunitx}
\usepackage{tabulary}
\usepackage{xcolor}
\hypersetup{
    colorlinks,
    linkcolor={red!50!black},
    citecolor={blue!50!black},
    urlcolor={blue!80!black}
}

\begin{document}
\title{Astrometric and photometric calibrators for the 4-m International Liquid Mirror Telescope}

\author[affil={1,2}, corresponding]{Naveen}{Dukiya}
\author[affil={1,3}]{Bhavya}{Ailawadhi}
\author[affil={4, 5}]{Talat}{Akhunov}
\author[affil={6}]{Ermanno}{Borra}
\author[affil={1,2}]{Monalisa}{Dubey}
\author[affil={7}]{Jiuyang}{Fu}
\author[affil={7}]{Baldeep}{Grewal}
\author[affil={7}]{Paul}{Hickson}
\author[affil={1}]{Brajesh}{Kumar}
\author[affil={1}]{Kuntal}{Misra}
\author[affil={1,3}]{Vibhore}{Negi}
\author[affil={1,8}]{Kumar}{Pranshu}
\author[affil={7}]{Ethen}{Sun}
\author[affil={9}]{Jean}{Surdej}
\affiliation[1]{Aryabhatta Research Institute of Observational sciencES (ARIES), Manora Peak, Nainital, 263001, India}
\affiliation[2]{Department of Applied Physics, Mahatma Jyotiba Phule Rohilkhand University, Bareilly, 243006, India}
\affiliation[3]{Department of Physics, Deen Dayal Upadhyaya Gorakhpur University, Gorakhpur, 273009, India}
\affiliation[4]{National University of Uzbekistan, Department of Astronomy and Astrophysics, 100174 Tashkent, Uzbekistan}
\affiliation[5]{ Ulugh Beg Astronomical Institute of the Uzbek Academy of Sciences, Astronomicheskaya 33, 100052 Tashkent, Uzbekistan}
\affiliation[6]{Department of Physics, Universit\'{e} Laval, 2325, rue de l'Universit\'{e}, Qu\'{e}bec, G1V 0A6, Canada}
\affiliation[7]{Department of Physics and Astronomy, University of British Columbia, 6224 Agricultural Road, Vancouver, BC V6T 1Z1, Canada}
\affiliation[8]{Department of Applied Optics and Photonics, University of Calcutta, Kolkata, 700106, India}
\affiliation[9]{Institute of Astrophysics and Geophysics, University of Li\`{e}ge, All\'{e}e du 6 Ao$\hat{\rm u}$t 19c, 4000 Li\`{e}ge, Belgium}

\correspondance{ndukiya@aries.res.in, ndookia@gmail.com}
\date{13th October 2020}
\maketitle


%

\begin{abstract}
The International Liquid Mirror Telescope (ILMT) is a 4-meter class survey telescope. It achieved its first light on 29$^{\rm th}$ April 2022 and is now undergoing the commissioning phase. It scans the sky in a fixed \ang{;22;} wide strip centred at the declination of $+$\ang{29;21;41.4} and works in \emph{Time Delay Integration (TDI)} mode. 
We present a full catalog of sources in the ILMT strip derived by crossmatching \textit{Gaia} DR3 with SDSS DR17 and PanSTARRS-1 (PS1) to supplement the catalog with apparent magnitudes of these sources in $g, r$, and $i$ filters. These sources can serve as astrometric calibrators. The release of Gaia DR3 provides synthetic photometry in popular broadband photometric systems, including the SDSS $g, r$, and $i$ bands for $\sim$220 million sources across the sky. We have used this synthetic photometry to verify our crossmatching performance and, in turn, create a subset of the catalog with accurate photometric measurements from two reliable sources.
\end{abstract}

\keywords{Liquid Mirror Telescope, Survey, Crossmatching, Astrometry, Photometry}

\section{Introduction}
\label{introduction}

Liquid mirror telescopes (LMTs) are a class of telescopes in which the primary mirror is made of a thin layer of liquid, usually mercury. \cite{gibson91} describes the working of liquid mirror telescopes where the liquid in a rotating container takes the shape of a paraboloid and acts as a mirror. Due to its working principle, LMTs can only perform zenithal observations. Despite their shortcomings, the high observing efficiency, optimal seeing conditions, best transparency, and less light pollution at the zenith make them very useful for conducting sky surveys. The 4m International Liquid Mirror Telescope (ILMT; \citealt{surdej2018}) is the newest of this class, situated at Devasthal, India. It is the first optical survey telescope in this part of the globe and performs observation in SDSS $g, r$, and $i$ bands. The ILMT performs zenithal observations in a \ang{;22;} wide strip centred at the declination of $+$\ang{29;21;41.4}. The CCD imager is operated in Time Delay Integration (TDI) mode to electronically track sources in order to compensate for the Earth's rotation, providing an effective integration time of 102s. The effect of star-trail curvature and differential drift is compensated by a 5-piece optical corrector system \citep{hickson_optical_corrector}. The ILMT observes the same strip of sky each night, making it particularly useful in the study of transients and variable sources. Also, images from the same patch of the sky can be co-added together to gaze deeper into the sky.

The images taken using the CCD are in pixel coordinates. In order to convert these coordinates to standard equatorial coordinates (e.g. J2000 or ICRS), it is essential to have a catalog of sources with precisely measured positions. \cite{dukiya_naveen} presented such a catalog derived by crossmatching Gaia EDR3 \citep{gaiaedr3} with Sloan Digital Sky Survey (SDSS) - DR17 \citep{sdssdr17} and The Panoramic Survey Telescope \& Rapid Response System (Pan-STARRS or PS1; \citealp{panstarrs_survey}).
The characteristics of the sources for astrometric calibration are extracted from \textit{Gaia} EDR3 as it provides a very precise measurement of astrometric properties such as RA ($\alpha$), Dec ($\delta$), parallax ({$\pi$}), and proper motions ($\mu_{\alpha^{*}}$ \& $\mu_{\delta}$).
The crossmatching was done in order to remove spurious sources and have $g, r$, and $i$ magnitudes of these sources. The release of Gaia DR3 brings access to synthetic photometry in multiple broad-band photometric systems derived from the low-resolution BR and RP spectra.
In this work, this information is used to investigate the general accuracy of the crossmatching results.
In addition, we have created a subset of this catalog where the crossmatching is also verified by photometry which can be used as potential photometric calibrators for the ILMT images. The methodology and results are discussed in Section \ref{sec:methods}. A summary of this work is presented in Section \ref{sec:summary}.

\section{Methodology} \label{sec:methods}
ESA's \textit{Gaia} satellite has recently had its third public data release \citep{gaia_mission, gaiadr3}. In addition to astrometry and broad-band photometry released as a part of Gaia EDR3, this data release brings flux-calibrated very low-resolution spectra in the 330nm - 1050nm range for sources having G < 17.65 from the BP and RP spectrographs. The internal calibration of these spectra is described in \cite{xp_spectra_internal_angeli} and \cite{xp_spectra_internal_carrasco}. Their external calibration with respect to a set of spectrophotometric standard stars \citep{gaia_spss_pancino} is described in \cite{xp_external_montegriffo}.
These externally calibrated spectra can be used to derive synthetic photometry in any passbands enclosed within this wavelength range using the framework described by \cite{bessel_murphy_2012}. These synthetic magnitudes are provided as part of the Gaia DR3 \citep{gaia_synth_phot} in popular broad-band photometric systems (including SDSS $g, r$ and $i$). These synthetic magnitudes are standardized against primary or secondary photometric standards of broadband photometric bandpasses (JKC, SDSS, PS1, etc.) to remove any systematic effects.

\cite{dukiya_naveen} presented a catalog suitable for the astrometric calibration of ILMT images. The algorithm used was position-based, and the nearest neighbour (up to 2 arcsec) from SDSS or PS1 was considered a crossmatch to the \textit{Gaia} object. An additional good-neighbour filter \citep{pineau} was applied to remove low-confidence crossmatches based on the constraint provided by the joint positional error of a source (see Section 3 of \citealp{dukiya_naveen} for details). In this work, we have updated the catalog with the data products provided by \textit{Gaia} DR3. This gives us another measurement of the $g, r$, and $i$ band magnitudes of some of these sources, which can be compared with the cataloged magnitude of crossmatched counterparts from SDSS and PS1 to gauge the accuracy of the crossmatch. 

\begin{figure}
    \centering
    \includegraphics[width=\textwidth]{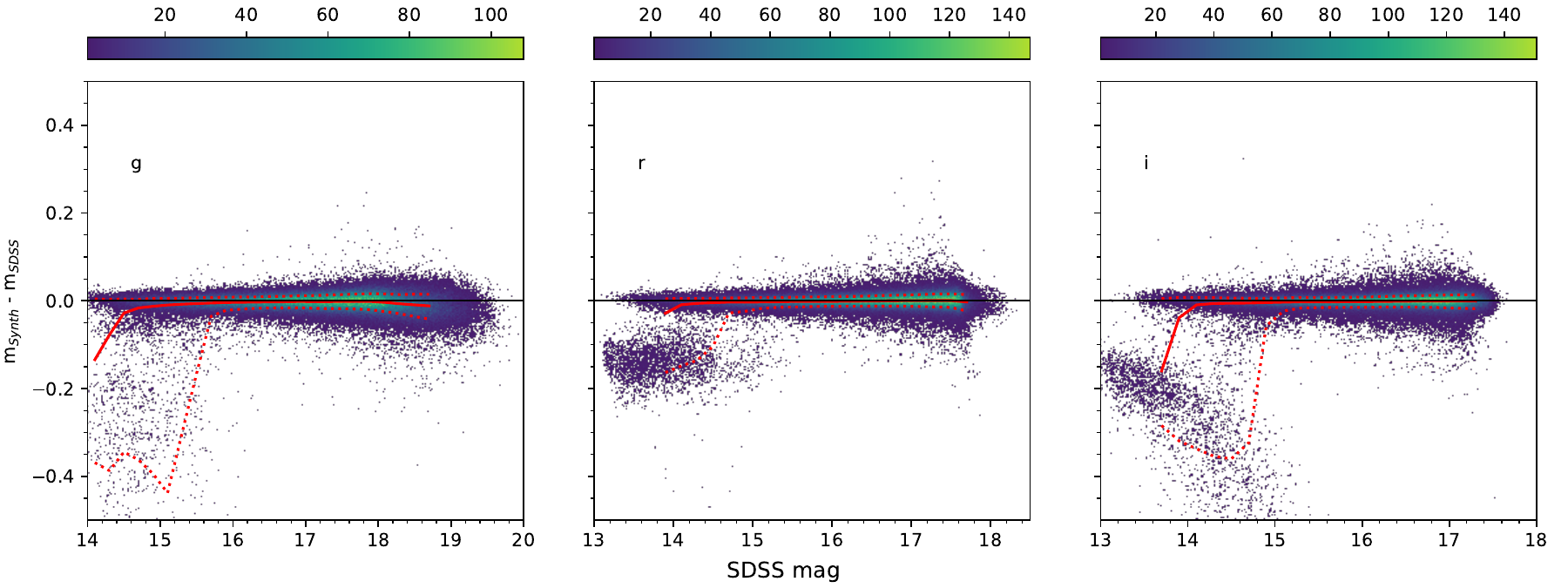}
    \begin{minipage}{12cm}
    \caption{Deviation between \textit{Gaia} DR3 synthetic magnitude and SDSS magnitude of the crossmatched counterparts. The red lines represent the median and 1-sigma scatter of the deviations. The thin black line represents zero-deviation. The colorbar represents the number of sources in one pixel of the image.} 
    \label{fig:sdss_uncleaned_deviation}
    \end{minipage}
\end{figure}

\autoref{fig:sdss_uncleaned_deviation} shows the deviation between the synthetic magnitude of sources and the catalog magnitude of the counterparts. For most sources, the deviation is much less, as indicated by the 1$\sigma$ lines. However, there are clear outliers in the data. To reduce the number of outliers, we consider the following three quality indicators. \\
1. \textbf{C$^*$} (or corrected BP and RP flux excess factor): The factor C represents the flux excess in $G_{BP}$ and $G_{RP}$ compared to the $G$ band. C$^*$ represents the deviation from expected flux excess at any given color (for a detailed description, see \citealt{gaia_phot_verification_riello}). We have chosen sources that have C$^* < \sigma$, where $\sigma$ represents the scatter in C$^*$ at a given magnitude.\\
2. \textbf{\textit{Gaia} Variability Flag}: \textit{Gaia} DR3 includes \texttt{phot\_variable\_flag} to indicate the variability of sources. Even though for the majority of sources, the status is unclear, we have removed the sources which are confirmed variables.\\
3. The synthetic photometry table includes a flag for every band, which is 1 if the color and magnitude of the source are within the range where standardization was performed. The flag is zero if standardization is extrapolated. We have only taken sources where this flag is 1.

\begin{figure}
    \centering
    \includegraphics[width=\textwidth]{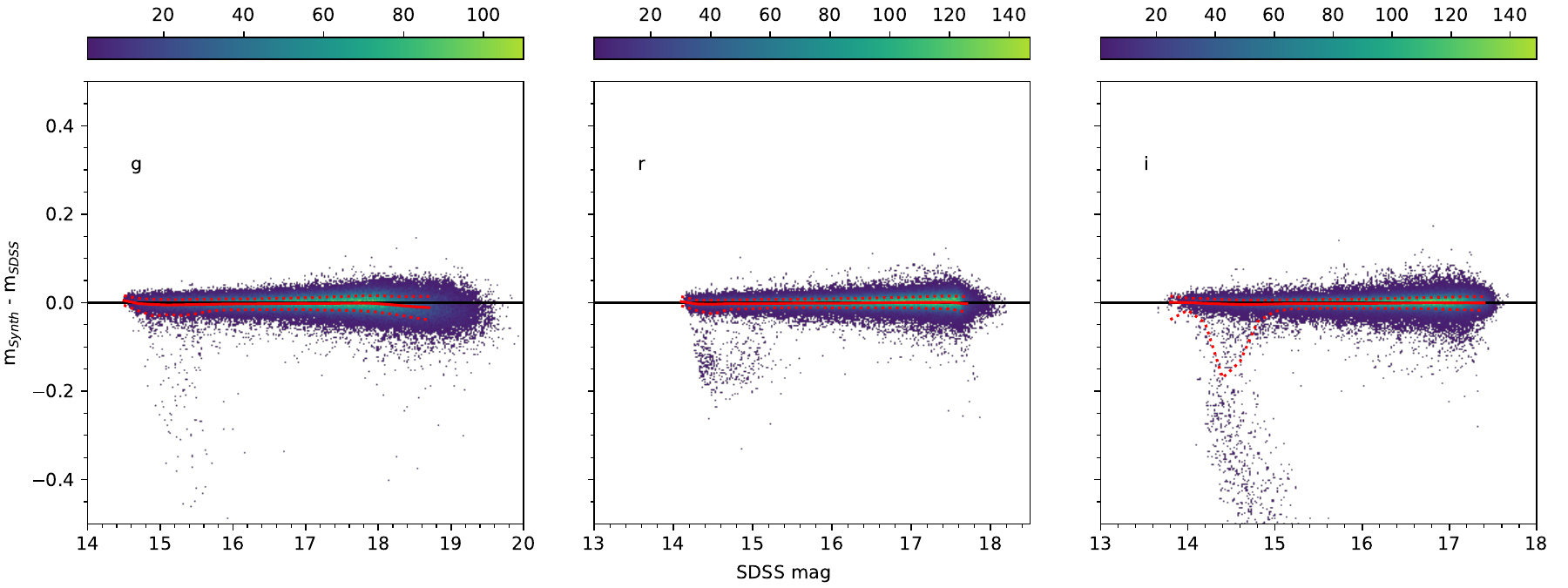}
    \begin{minipage}{12cm}
    \caption{Same as \autoref{fig:sdss_uncleaned_deviation}, but confirmed variable sources and sources with unreliable photometry have been removed.}
    \label{fig:sdss_cleaned_deviation}
    \end{minipage}
\end{figure}

After applying the above-mentioned quality cuts, the cleaned sample is plotted in \autoref{fig:sdss_cleaned_deviation}. The statistics of these sources are tabulated in \autoref{tab:objsummary}.

\begin{table}[h]
    \centering
    \caption{Summary of the number of objects at different stages of this work.}
    \label{tab:objsummary}
    \vspace{0.5cm}
    \begin{tabulary}{\linewidth}{|c|C|C|C|}
    \hline
        Catalog & Total objects crossmatched & Crossmatched objects with \textit{Gaia} synthetic photometry & Objects in the cleaned sample \\ \hline
        SDSS &
        0.89 million&
        0.19 million&
        0.17 million \\
        PS1 &
        3.04 million&
        0.38 million&
        0.31 million\\ \hline
    \end{tabulary}
\end{table}


It is clear that a lot of initial outliers can be attributed to the above three factors. Still, some outliers are left, which may arise due to the following reasons - 
1. Some sources may be variable but not flagged as confirmed variables by \textit{Gaia}.
2. Imperfect calibration of the BP and RP spectra or imperfect standardization of the derived magnitudes.
3. Misidentified counterpart during crossmatching. It is more likely to occur in crowded fields.\\
We note that the outliers seem to be systematic in nature. We expect outliers due to 1 and 3 to be randomly distributed around zero. Additionally, the fields covered by SDSS are mostly sparse as SDSS does not cover the low galactic latitudes. Therefore, the probability of a misidentified counterpart is very low. To further investigate this,  we select only those source-counterpart pairs that have a small angular separation and a low probability of being a false match. This, however, does not reduce the number of outliers.
 
If we work under the assumption that any inconsistencies between the two photometric measurements are caused by mismatched counterparts, we can derive a lower limit for the accuracy of the cross-match results. If we allow for a deviation of 3$\sigma$ ($\sigma$ being the combined error $\sqrt{\sigma_1^2 + \sigma_2^2}$ of both measurements) in all three bands, we derive a lower limit of 70\% for this dataset. This lower limit increases to 86\% for 5$\sigma$ deviation. We expect the actual accuracy to be significantly higher than this as a significant portion of outliers show systematic error, as discussed above. Since the $i$ band has the most outliers, only considering $g$ \& $r$ bands gives us a lower limit of 78\% and 91\% for 3$\sigma$ and 5$\sigma$ deviations, respectively. Even this estimate is conservative as there may be many variable sources and systematic effects at play.

\begin{figure}
    \centering
    \includegraphics[width=\textwidth]{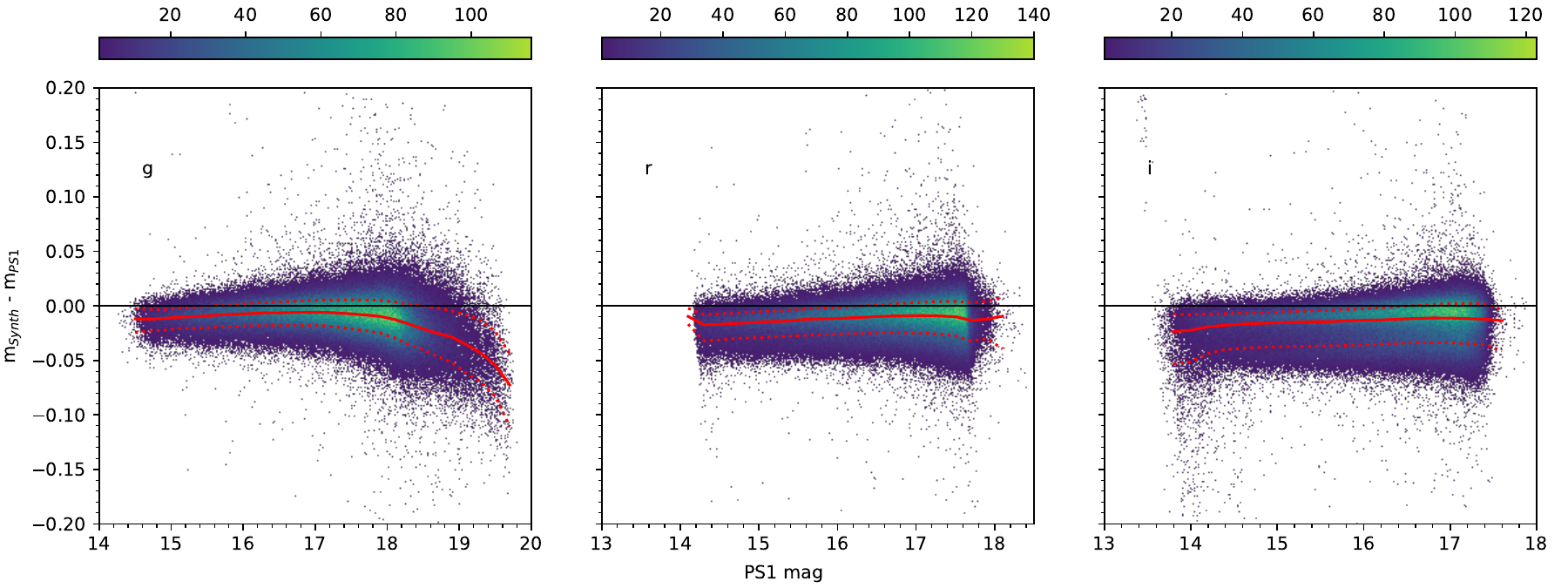}
    \begin{minipage}{12cm}
    \caption{Deviation between \textit{Gaia} DR3 synthetic magnitude and PS1 magnitude of the crossmatched counterpart. The meaning of the symbols and colorbar is the same as \autoref{fig:sdss_uncleaned_deviation}.}
    \label{fig:ps1_cleaned_deviation}
    \end{minipage}
\end{figure}

In the low galactic latitude fields, SDSS has limited coverage. To identify astrometric calibrators in this region, \cite{dukiya_naveen} had crossmatched \textit{Gaia} EDR3 with PS1. \autoref{fig:ps1_cleaned_deviation} shows the deviations for the cleaned PS1 sample. The magnitudes in the PS1 bandpasses were derived using \texttt{GaiaXPy}, which makes use of the calibrated BP and RP spectra as the standardized PS1 photometry is not a part of the final data products in \textit{Gaia} DR3. Unlike the SDSS sample, we see a significant systematic trend in all the sources. \cite{gaia_synth_phot} notes that the standardization for PS1 bandpasses is not as thorough as SDSS. This could be a likely reason for the systematic trend noticed and will require further investigation.

\section{Summary} \label{sec:summary}
In this work, we have expanded upon the work presented in \cite{dukiya_naveen}. We have included \textit{Gaia} DR3 data-products and Gaia Synthetic photometry into the \textit{Astrometric and Photometric Calibration Catalog}. We have characterized the deviations between the \textit{Gaia} synthetic magnitudes and SDSS catalog magnitudes of the counterparts using a cleaned sample of sources that have reliable synthetic photometry. This has allowed us to put a lower limit on the accuracy of \textit{Gaia}-SDSS crossmatching. It should be noted that the crossmatching accuracy will likely be lower for denser fields. In future, detailed characterization of the PS1 crossmatching will shed light on this. 

This catalog contains 0.24 million sources fainter than the saturation limit of the ILMT in the $i$ band ($\sim$ 16.5 mag) covering the entire ILMT strip. A subset of these sources in which the magnitudes from both measurements agree to a desired level (say 1$\sigma$) can be used to estimate the photometric zero-point of the frames observed by the ILMT. In addition, this catalog can be useful in identifying long-term variability or outbursts in astrophysical sources as the data from all the different surveys span decades, which can be compared with the ILMT images with precise astrometry and photometry. Finally, this dataset can be used to identify potential targets-of-interests (e.g. AGNs, white dwarfs, and variable stars) for follow-up by locating them in color-color and color-magnitude diagrams.

\begin{acknowledgments}
We thank the anonymous referee for the comments that helped improve this paper. The 4m International Liquid Mirror Telescope (ILMT) project results from a collaboration between the Institute of Astrophysics and Geophysics (University of Li\`{e}ge, Belgium), the Universities of British Columbia, Laval, Montreal, Toronto, Victoria and York University, and Aryabhatta Research Institute of observational sciencES (ARIES, India). The authors thank Hitesh Kumar, Himanshu Rawat, Khushal Singh and other observing staff for their assistance at the 4m ILMT. The team acknowledges the contributions of ARIES's past and present scientific, engineering and administrative members in the realisation of the ILMT project. JS wishes to thank Service Public Wallonie, F.R.S.-FNRS (Belgium) and the University of Li\`{e}ge, Belgium, for funding the construction of the ILMT. PH acknowledges financial support from the Natural Sciences and Engineering Research Council of Canada, RGPIN-2019-04369. PH and JS thank ARIES for hospitality during their visits to Devasthal. B.A. acknowledges the Council of Scientific $\&$ Industrial Research (CSIR) fellowship award (09/948(0005)/2020-EMR-I) for this work. M.D. acknowledges Innovation in Science Pursuit for Inspired Research (INSPIRE) fellowship award (DST/INSPIRE Fellowship/2020/IF200251) for this work. T.A. thanks Ministry of Higher Education, Science and Innovations of Uzbekistan (grant FZ-20200929344).
\end{acknowledgments}

\begin{furtherinformation}

\begin{orcids}
\orcid{0000-0002-0394-6745}{Naveen}{Dukiya}
\orcid{0000-0001-7225-2475}{Brajesh}{Kumar}
\orcid{0000-0003-1637-267X}{Kuntal}{Misra}
\orcid{0000-0001-5824-1040}{Vibhore}{Negi}
\orcid{0000-0002-7005-1976}{Jean}{Surdej}
\end{orcids}

\begin{authorcontributions}
This work results from a long-term collaboration to which all authors have made significant contributions.
\end{authorcontributions}

\begin{conflictsofinterest}
The authors declare no conflict of interest.
\end{conflictsofinterest}

\end{furtherinformation}

\bibliographystyle{bullsrsl2-en}

\bibliography{extra}

\end{document}